# On the Best Bandstructure for Thermoelectric Performance

Changwook Jeong, Raseong Kim, and Mark Lundstrom

Network for Computational Nanotechnology

Birck Nanotechnology Center

Purdue University

West Lafayette, Indiana, 47907

**Abstract**- The conventional understanding that a bandstructure that produces a Dirac delta function transport distribution (or transmission in the Landauer framework) maximizes the thermoelectric figure of merit, *ZT,* is revisited.  Thermoelectric (TE) performance is evaluated using a simple tight binding (TB) model for electron dispersion and three different scattering models: 1) a constant scattering time, 2) a constant mean-free-path, and 3) a scattering rate proportional to the density-of-states. We found that a δ-function transmission never produces the maximum *ZT*. The best bandstructure for maximizing *ZT* depends on the scattering physics. These results demonstrate that the selection of bandstructure to maximize TE performance is more complex than previously thought and that a high density-of-states near the band edge does not necessarily improve TE performance.

1. Introduction

Increasing the thermoelectric figure of merit (FOM), $ZT = S^2\sigma T_L/\kappa$, involves increasing the power factor ($PF = S^2\sigma$ where S is the Seebeck coefficient and $\sigma$ is the electrical conductivity) and/or decreasing the thermal conductivity, $\kappa$, which is the sum of the electronic thermal conductivity at zero current, $\kappa_{el}$, and the lattice thermal conductivity, $\kappa_{ph}$. Recently, there has been much success in enhancing TE performance by reducing $\kappa_{ph}$ without significantly degrading $\sigma$ [1-3]. This leads to the question of whether the numerator of the FOM can be enhanced. Mahan and Sofo addressed this important problem and concluded that the best bandstructure is one that produces a δ-function transport distribution [4]. (The transport distribution in conventional thermoelectrics is equivalent to the transmission in the Landauer framework [2]). As Mahan and Sofo showed, the δ-function transmission makes $\kappa_{el} = 0$ and the FOM equal to $\kappa_0/\kappa_{ph}$, where $\kappa_0$ is the electronic thermal conductivity for no voltage gradient. We will show that the δ-function transmission does, indeed, produce $\kappa_{el} = 0$ and a FOM equal to $\kappa_0/\kappa_{ph}$, but it does not maximize the power factor or, except for the case of $\kappa_{ph} = 0$, the FOM. Although we assume a specific shape for the electron dispersion and consider only three specific scattering models, we expect the results to apply more generally. This work contributes to the understanding of how to engineer bandstructures to maximize TE performance, a question that was first addressed by Mahan and Sofo and that is still very important [4].

2. Approach

We begin with a brief review of the Landauer approach to thermoelectric transport as presented in [5, 6]. Although this approach is more general, in this paper we restrict our attention to three-dimensional, diffusive samples. Accordingly, the expression for the differential conductivity becomes

$$\sigma'(E) = \left(\frac{2q^2}{h}\right)\frac{M_{el}(E)}{A}\lambda_{el}(E)\left(-\frac{\partial f_0}{\partial E}\right) = q^2 \Sigma(E)\left(-\frac{\partial f_0}{\partial E}\right), \quad (1)$$

where $2q^2/h$ is the quantum of electrical conductance, $M_{el}(E)$ the number of conduction channels at $E$, $A$ the cross-sectional area, $\lambda_{el}(E)$ the mean-free-path (MFP) for backscattering, and $f_0$ the Fermi-Dirac distribution. Conventionally, thermoelectric parameters are derived by solving the diffusive Boltzmann Transport Equation (BTE), and the results are expressed in terms of $\Sigma(E) = (2/h)(M(E)/A)\lambda_{el}(E)$, the so-called transport distribution [4]. The BTE and Landauer approaches are mathematically equivalent in the diffusive limit, but the Landauer approach provides a convenient way to treat ballistic and quasi-ballistic transport as well [5].

In terms of the differential electrical conductivity, the transport coefficients can be written as [4-6],

$$\sigma = I_0, \quad \sigma S = (k_B/q)I_1, \quad \kappa_0 = (k_B/q)^2 T_L I_2, \quad (2a)$$

with the integrals $I_n$ being defined as:

$$I_n = \int_{-\infty}^{+\infty} dE (E - E_F)^n \sigma'(E), \quad (2b)$$

The thermoelectric figure of merit, $ZT$, is

$$ZT = \frac{S^2 \sigma T_L}{\kappa_{el} + \kappa_{ph}}, \quad (3a)$$

where

$$\kappa_{el} = \kappa_0 - S^2 \sigma T_L. \quad (3b).$$

From the definition of $\kappa_{el}$ in terms of the integrals, $I_n$, it is readily seen that $\kappa_{el} = 0$ when the differential conductivity (or transport distribution or transmission) is a $\delta$-function. In that case, $\kappa_0 = S^2 \sigma T_L$ and $ZT \to \infty$, as noted in [4]. We are concerned in this paper with the more typical case where $\kappa_{ph} > \kappa_{el}$ and ask: What bandstructure maximizes the power factor, $S^2 \sigma$?

We begin by reviewing the approach to evaluate the number of conducting channels and mean-free-path for backscattering. From a given electron dispersion, $M_{el}(E)$ can be obtained in a computationally simple way by counting the bands that cross the energy of interest [5]. In this paper, a simple, nearest neighbor TB model is used to describe the electron dispersion,

$$E_k = E_C + 2t_0(1-\cos k_x a) + 2t_0(1-\cos k_y a) + 2t_0(1-\cos k_z a), \quad (4)$$

where $t_0 = \hbar^2/2m_e a^2$ with $a$ and $m_e$ being the lattice constant and the effective electron mass, respectively. The bandwidth (BW) of the electron dispersion is $12t_0$. We change the BW by varying $m_e$ while assuming $a = 5 \times 10^{-10}$ m.

Three different, simple models are used for the mean-free-path for backscattering. In the first model, $\lambda_{el}(E)$ is assumed to be energy-independent, $\lambda_{el}(E) = \lambda_0$. In the second model, we assume that the momentum relaxation time, $\tau(E) = \tau_0$ is constant, which has been typically used [7-10]. (The relaxation time and mean-free-path for backscattering are related by $\lambda_{el}(E) = (4/3)\upsilon(E)\tau(E)$ [5].) Finally, we also consider an energy-dependent MFP in which, the scattering rate is assumed to be proportional to the density of states, $\tau_{el}^{-1}(E) = C_{el} D(E)$ with $C_{el}$ being a constant and $D(E)$ the density-of-states (*DOS*). This assumption is reasonable when scattering occurs due to an acoustic deformation potential or to a strongly screened coulomb potential. In practice, scattering is a complicated problem, but these three models will give some sense of how scattering affects thermoelectric performance. To calibrate these models, we begin with a large BW dispersion ($m_e = m_0$) and set $\lambda_0 = 10$ nm. For the second model, we select $\tau_0$ to give the same average MFP, and for the third model, we select $C_{el}$ to give 10 nm. As we vary the BW, these parameters are not changed.

## 3. Results

In this section, the thermoelectric coefficients will be evaluated as the bandwidth of the dispersion is varied. For each of the three scattering models considered, the results show that a δ-function transmission is not the best bandstructure for maximum TE performance.

In Fig. 1a, we plot $\kappa_{el}$, as a function of the electron effective mass – i.e., the BW of the electron dispersion. For each value of the electron effective mass, we found the optimal location of the Fermi level to maximize the *PF*, so the results for each effective mass in Fig. 1a corresponds to a different location of the Fermi level. Figure 1a shows that as the effective mass increases (BW decreases) and the transmission approaches a δ-function, $\kappa_{el}$ approaches zero, as noted by Mahan and Sofo [4].

In Fig. 1b, we plot the dimensionless Lorenz number $L \equiv (q/k_B)^2 (\kappa_{el}/\sigma T_L)$ vs. Fermi level for three different dispersions. For a parabolic energy band, the Lorenz number is

$$L = \left(\frac{q}{k_B}\right)^2 \frac{\kappa_{el}}{\sigma T_L} = (r+2) \tag{5a}$$

for Maxwell-Boltzmann statistics and

$$L = \left(\frac{q}{k_B}\right)^2 \frac{\kappa_{el}}{\sigma T_L} = \frac{(r+3)(r+2)\mathscr{F}_{r+2}(\eta_F)\mathscr{F}_r(\eta_F) - (r+2)^2 \mathscr{F}_{r+1}^2(\eta_F)}{\mathscr{F}_r^2(\eta_F)} \tag{5b}$$

for Fermi-Dirac statistics. In Eq. (5), $\mathscr{F}_j(\eta_F)$ is the Fermi-Dirac integral of order $j$ defined as $\mathscr{F}_j(\eta_F) = 1/\Gamma(j+1) \times \int_0^\infty dx\, x^j / (1 + e^{x-\eta_F})$, $\eta_F$ the position of $E_F$ relative to the band edge, and $r$ is a characteristic exponent that describes scattering, $\lambda_{el} = \lambda_0 (E/k_B T_L)^r$. As shown Fig. 1b, a typical behavior of $L(E_F)$ is observed for the large BW band case, but as the BW decreases, $L$ decreases, and $L \to 0$ for the near δ-function transmission. Figure 1b also shows that the scattering model has little effect on $L$.

In Fig. 1c, the quantity, $S^2 \sigma T_L / \kappa_0$ is plotted as a function of the electron effective mass – i.e., BW of the electron dispersion. The location of the Fermi level was selected to maximize the PF. The numerator of the thermoelectric FOM is $S^2 \sigma T_L$, and the two electronic thermal conductivities are related by $\kappa_{el} = \kappa_0 - S^2 \sigma T_L$. For a δ-function transmission, $\kappa_{el} = 0$ and $S^2 \sigma T_L$ approaches $\kappa_0$. Figure 1c shows that the ratio $S^2 \sigma T_L / \kappa_0$ approaches one as the BW decreases - regardless of the scattering model assumed. Accordingly, the thermoelectric FOM approaches $\kappa_0/\kappa_{ph}$ as the bandwidth approaches zero. These observations are consistent with the conclusions of Mahan and Sofo.

Finally, we turn to the question of which bandwidth produces the best FOM. In Fig. 2, we plot $S^2\sigma T_L$ and $\kappa_0$ vs. the bandwidth of the electron dispersion with the Fermi level selected to maximize the PF. In each case, we see that $S^2\sigma T_L \to \kappa_0$ as the bandwidth approaches zero. Note, however, that the results depend on how scattering is modeled. For the constant MFP or constant scattering time, the maximum $S^2\sigma T_L$ (and therefore the maximum FOM) occurs for a moderate bandwidth – not for a zero bandwidth. For a scattering rate proportional to the density-of-states, there is no optimum bandwidth. The broader the bandwidth, the higher the FOM. When the bandwidth is very small (effective mass large), the FOM approaches $\kappa_0/\kappa_{ph}$ but this is the maximum FOM only when $\kappa_{ph} = 0$.

As discussed for Eq. (3), our calculations agree with Mahan and Sofo's conclusions that $\kappa_{el} \to 0$ and the FOM approaches $\kappa_0/\kappa_{ph}$ as the transmission approaches a δ-function. Except for the case of $\kappa_{ph} = 0$, however, we find that a finite BW gives a higher FOM. This occurs because the upper bound of FOM, $\kappa_0/\kappa_{ph}$, also depends on the transmission, and the largest value of $\kappa_0/\kappa_{ph}$ occurs for the finite BW – not for a zero bandwidth. Although the finite BW gives $\kappa_{el} > 0$, this is more compensated for by the resulting increase in the PF. For a constant MFP or a constant scattering time, we find that a small, but finite BW is best, but for a scattering rate proportional to the density-of-states, a very broad band is best.

**4. Discussion**

In this section, we discuss why a bandstructure that produces a δ-function-like transport distribution (or transmission in the Landauer framework) does not necessarily maximize the FOM.

Figure 3a and 3b shows the Seebeck coefficient ($S_{opt}$) and the electrical conductivity ($\sigma_{opt}$) as a function of the bandwidth of the electron dispersion for the optimal location of the Fermi level to maximize the *PF*. In Fig. 3a and 3b, we observe that $\sigma_{opt}$ depends more sensitively on both the bandwidth and the scattering models than $S_{opt}$, indicating that the best bandstructure for the maximum $S^2 \sigma T_L$ is mostly determined by $\sigma_{opt}$. As seen in Fig. 3a, each of the three scattering models produced similar results for $S_{opt}$; a zero bandwidth gives only ~30% increase in $S_{opt}$ comparing to the $S_{opt}$ for a finite bandwidth. Note that the non-monotonic behavior of $S_{opt}$ for a constant MFP occurs because of non-parabolic dispersion relation, and the behaviors disappears when we include bandstructure dependent scattering – i.e., a constant scattering time or a scattering rate proportional to the DOS. In contrast, a zero bandwidth band gives rise to about ten-fold decrease in $\sigma_{opt}$ over the optimal bandwidth case (Fig. 3b). The profile for a constant scattering time or a constant MFP is very different from the scattering rate proportional to the DOS case in that the maximum $\sigma_{opt}$ appears at a moderate BW.

Figure 3b show that the best bandstructure is mostly determined by $\sigma_{opt}$, which is proportional to $(2q^2/h)\langle \bar{T}_{el} \rangle$. The average transmission, $\langle \bar{T}_{el} \rangle$, is defined as $\langle \bar{T}_{el} \rangle \equiv \int \bar{T}_{el}(E)(-\partial f_0/\partial E) dE$ with the transmission, $\bar{T}_{el}(E)$ being $\lambda_{el}(E) M_{el}(E)$, so to understand how the BW affect $\sigma_{opt}$, we need to understand how BW affects $M_{el}(E)$ and $\lambda_{el}(E)$.

In Fig. 4a, the normalized number of conducting channels, $M_{el}(E)$, is plotted as a function of energy for three different BWs, $m_0$, $10m_0$, and $100m_0$. Note that in each case $M_{el}(E)$ displays the identical peak value. It can be seen in the insert that the BW of $M_{el}(E)$ approached zero for the high effective mass dispersion ($100m_0$). For a constant MFP, the behavior of $\sigma_{opt}$ (i.e., $\langle \bar{T}_{el} \rangle$) in Fig. 3b is attributed to the BW of the $M_{el}(E)$. We found that the average transmission, $\langle \bar{T}_{el} \rangle$, increases with smaller BW/heavier effective mass until the BW of the dispersion is equal to the width of the Fermi "window function," $(-\partial f_0/\partial E)$, after which it decreases. Consequently, the maximum $\sigma_{opt}$ occurs for a moderate BW transmission with an effective mass of $6m_0$. The maximum $S^2\sigma T_L$, however, occur for a slightly larger effective mass ($10m_0$), which corresponds to a bandwidth of ~ $7k_B T_L$ because of the interplay between $\sigma_{opt}$ and $S_{opt}$ as seen in Fig. 3a and 3b. For this reason, we observe an optimum BW for maximum PF when a constant MFP is considered.

As seen in Fig. 4b, the MFP plays a critical role for a scattering rate proportional to the DOS, in contrast to the constant MFP case where only the BW of the $M_{el}(E)$ is important in determining $\langle \bar{T}_{el} \rangle$. The average MFP rapidly decreases with the bandwidth/effective mass because the small BW leads to short scattering time as well as small group velocity. Thus, a zero-bandwidth band is worst. For a constant scattering time, both the BW of the $M_{el}(E)$ and the MFP are important because the bandstructure affects the MFP only through the group velocity. As a result, we found that the optimal BW for maximum PF occurs at ~2×7$k_B T_L$ (i.e., $m_e = 5m_0$), which is smaller than the optimum BW for a scattering rate proportional to the DOS, but larger than the optimum BW for a constant MFP. Although the scattering models affect the optimum BW of transmission for maximum FOM, $\kappa_{el}$ always approaches zero (i.e., $S^2 \sigma T_L$ approaches $\kappa_0$) as the bandwidth approaches zero, regardless of the scattering model (Fig. 1). As noted by Mahan and Sofo [4], this occurs because $\kappa_{el}$ is proportional to the variance of the transmission that is almost independent of the scattering model considered.

In this section, we discussed the factors that determine optimum BW for maximum TE performance. For a constant MFP, the BW of the electron dispersion determines the optimum BW, whereas the MFP is more important than the BW for a scattering rate proportional to the DOS. For a constant scattering time, the optimum BW is determined by both the BW and the MFP.

**5. Conclusions.**

We revisited the conventional understanding that a bandstructure that produces a δ-function transmission maximizes the figure of merit, *ZT*. Our calculations agree with Mahan and Sofo's conclusions that $\kappa_{el} \to 0$ and that the FOM approaches $\kappa_0 / \kappa_{ph}$ as the transmission approaches a δ-function. We found, however, that the maximum *ZT* cannot be achieved with the δ-function transmission except for the case of $\kappa_{ph} = 0$ and that the best bandstructure for maximizing *ZT* depends on the scattering physics. This occurs because the upper bound of *ZT*, $\kappa_0/\kappa_{ph}$, also depends on the transmission and the δ-function transmission never maximize the $\kappa_0/\kappa_{ph}$. The finite BW gives $\kappa_{el} > 0$, but this is more compensated for by the resulting increase in the PF. For a constant MFP or a constant scattering time, we found that a moderate BW is best, but for a scattering rate proportional to the density-of-states, a very broad band is best. In this paper, we considered only a simple dispersion with different bandwidth, but the results provide motivation to re-examine TE performance of more complex bandstructures such as composite energy bands.

**ACKNOWLEDGEMENTS**

This work was supported by the Focus Center Research Program (FCRP-MSD). Computational resources were provided by the Network for Computational Nanotechnology (National Science Foundation under Cooperative Agreement No. EEC-0634750).

**Figures**

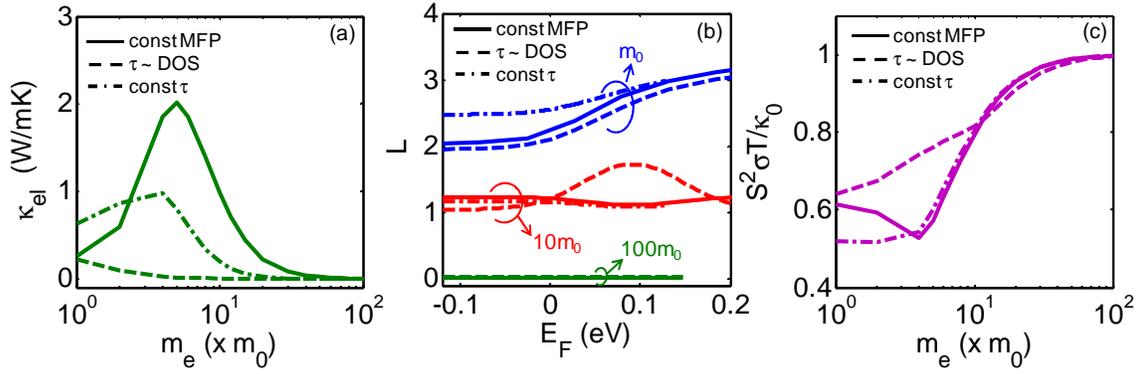

Figure 1. (a) The electronic thermal conductivity at zero current, $\kappa_{el}$, vs. the electron effective mass. (b) the dimensionless Lorenz number, $L \equiv (q/k_B)^2 (\kappa_{el}/\sigma T_L)$, vs. Fermi level. (c) $S^2 \sigma T_L / \kappa_0$ as a function of the electron effective mass. $S^2 \sigma T_L$ is the numerator of the $ZT$, and $\kappa_0$ the electronic thermal conductivity at zero voltage gradient.

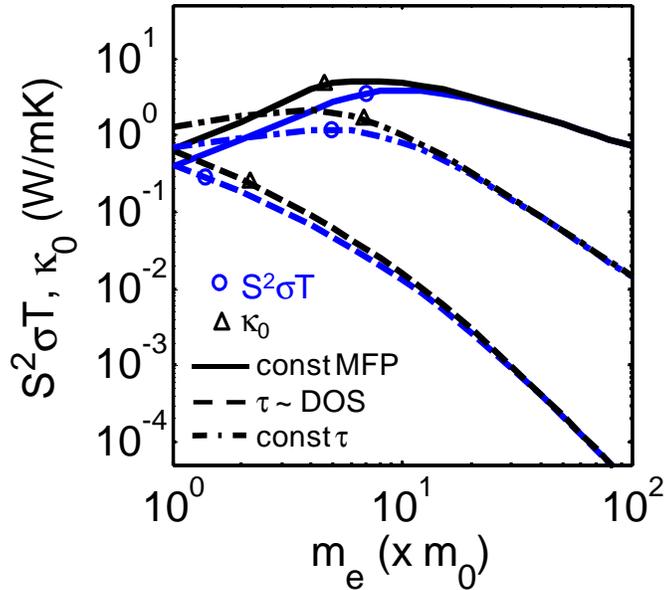

Figure 2. $S^2 \sigma T_L$ and $\kappa_0$ as a function of the electron effective mass.

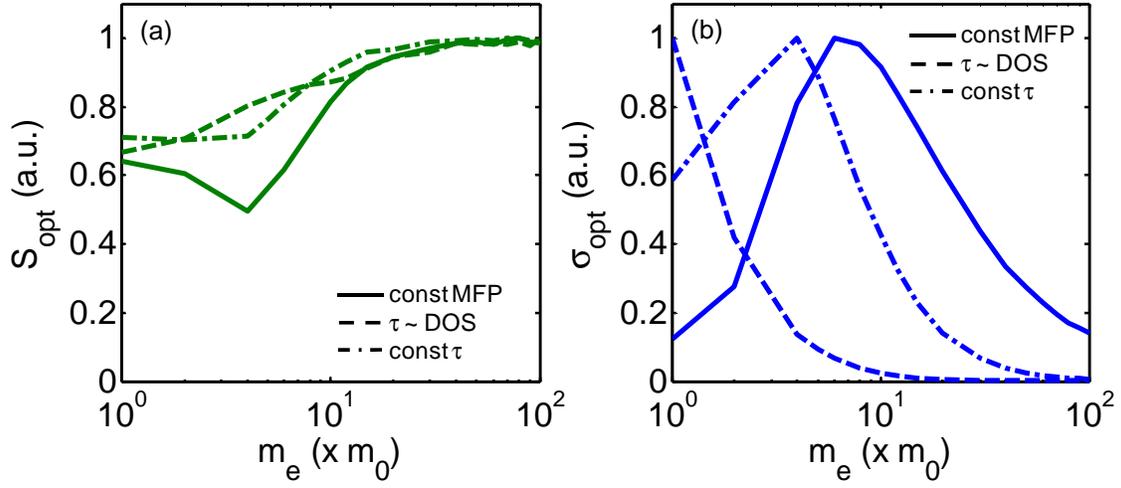

Figure 3. For the optimal location of the Fermi level to maximize the *PF*, (a) the Seebeck coefficient ($S_{opt}$) as a function of the electron effective mass. (b) the electrical conductivity ($\sigma_{opt}$) as a function of the electron effective mass.

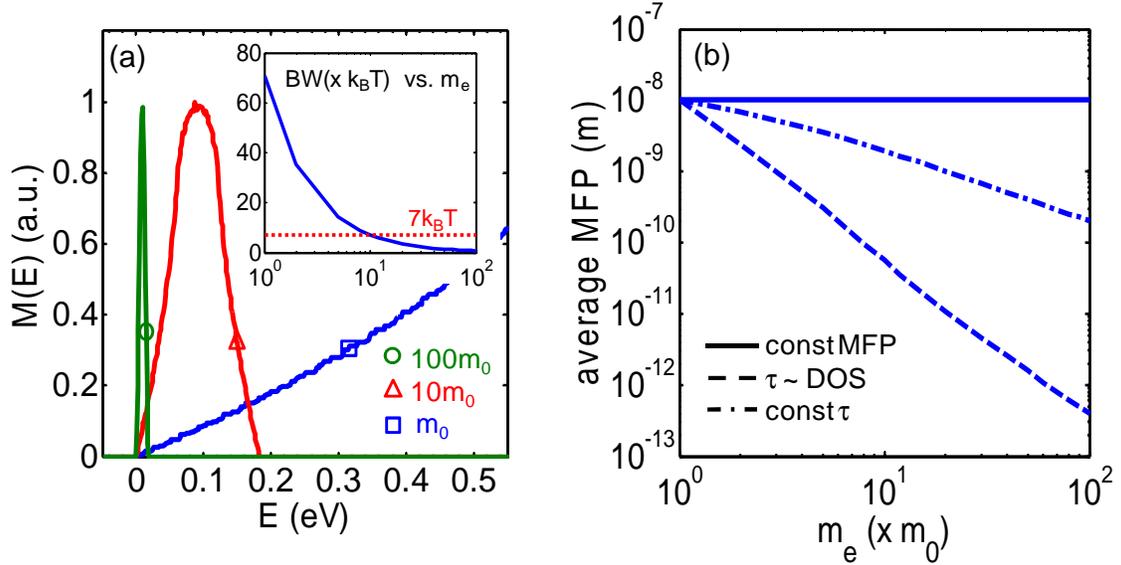

Figure 4. (a) Normalized number of conducting channels, $M_{el}(E)$, as a function of energy for three different BWs (three different effective mass: $m_0$, $10m_0$, $100m_0$). A normalization factor is $2.55 \times 10^{18}$ /m$^2$ ~ $0.6 N_A$ where the number of atoms in the cross-sectional area, $N_A = 1$ atoms/$a^2 = 4 \times 10^{18}$ atoms/m$^2$. Inset of (a): the bandwidth of the electron dispersion in the units of $k_B T_L$ as a function of the electron effective mass. The horizontal dotted line indicates $7 k_B T_L$. (b) average mean-free-path (MFP) as a function of the electron effective mass.